# A Secure Deep Probabilistic Dynamic Thermal Line Rating Prediction

N. Safari, *Member, IEEE*, S.M. Mazhari, *Member, IEEE*, C.Y. Chung, *Fellow, IEEE*, and S.B. Ko, *Senior Member*, *IEEE*

*Abstract*— **Accurate short-term prediction of overhead line (OHL) transmission ampacity can directly affect the efficiency of power system operation and planning. Any overestimation of the dynamic thermal line rating (DTLR) can lead to lifetime degradation and failure of OHLs, safety hazards, etc. This paper presents a secure yet sharp probabilistic prediction model for the hour-ahead forecasting of the DTLR. The security of the proposed DTLR limits the frequency of DTLR prediction exceeding the actual DTLR. The model is based on an augmented deep learning architecture that makes use of a wide range of predictors, including historical climatology data and latent variables obtained during DTLR calculation. Furthermore, by introducing a customized cost function, the deep neural network is trained to consider the DTLR security based on the required probability of exceedance while minimizing deviations of the predicted DTLRs from the actual values. The proposed probabilistic DTLR is developed and verified using recorded experimental data. The simulation results validate the superiority of the proposed DTLR compared to state-of-the-art prediction models using well-known evaluation metrics.**

## I. INTRODUCTION

THERMAL line rating (TLR) is the primary culprit limiting the current carrying capability of the overhead line (OHL) [1]. Both IEEE [2] and CIGRE [3] have put forward TLR calculation methods. Although these approaches provide very similar results, the one proposed by IEEE is simpler and easier to use [4]. The TLR of OHL is a weather-dependent variable and conventionally calculated using the heat balance equation under the worst-case weather scenarios [5]. Despite its conservativeness, in some cases this static TLR (STLR) might exceed the real OHL thermal constraint. Consequently, the OHL might be exposed to damage due to the lack of TLR monitoring.

To overcome the shortcomings of the STLR, dynamic TLR (DTLR) is proposed in which the thermal condition of OHL can be monitored. Ergo, DTLR unlocks the additional capacity headroom of current OHLs in a secure way, thereby addressing network congestion and postponing/eliminating the need for transmission expansion [6]. As a significant additional benefit, DTLR facilitates the delivery of highly variable and uncertain power from renewable energy systems (RESs) to the end-user due to perceptible correlation between RES generation and the additional capacity provided by the DTLR. For such reasons, DTLR has recently grasped the attention of governments and transmission companies and is considered as an enabling tool for enhancing the penetration of RESs [7]–[9]. DTLR applications and technologies are comprehensively reviewed in [10]-[11].

The DTLR is a function of several climatology variables, such as wind speed, wind direction, etc. [2]; therefore, its value for upcoming hours needs to be forecasted. The DTLR forecast can be employed in various power system problems, such as unit commitment, economic dispatch, optimal power flow, etc., [5], [12]. In this respect, specialized research communities have devoted momentous efforts to develop DTLR monitoring and prediction models [9]–[22]. Because DTLR monitoring and prediction may not be feasible through the entire OHL, critical spans are identified for this purpose. In [23]–[25], heuristic approaches are brought forward to identify the number and locations of the monitoring stations required to make the OHL fully observable from the TLR perspective.

Relying on the literature, DTLR prediction has been interpreted with two different viewpoints. In one group of studies, researchers predict the maximum allowable current at the OHL thermal limit [12]–[16]. In contrast, others consider DTLR prediction to refer to estimating the future value of an OHL temperature provided that it carries a certain amount of current [19]–[20]. While both perspectives bring about intriguing advancements to the field and provide insightful information about OHL thermal constraints, this paper zeroes in on DTLR prediction based on the first definition.

DTLR methods can be broadly divided into direct [12]–[18], indirect [19], and hybrid [20]-[22] approaches. In direct DTLR calculation, the required values are calculated based on the straightforward computation of the conductor's maximum allowable current while taking the maximum permissible

The views expressed in this paper are those of the authors and do not represent the view of SaskPower. This paper was prepared when N. Safari was with the Department of Electrical and Computer Engineering, University of Saskatchewan, Saskatoon, SK, S7N 5A9, Canada.

The work was supported in part by the Natural Sciences and Engineering Research Council (NSERC) of Canada and the Saskatchewan Power Corporation (SaskPower).

N. Safari is with Grid Operations Support, SaskPower, Regina, SK, S4P 0S1, Canada (e-mail: n.safari@usask.ca).

S.M. Mazhari, C.Y. Chung, and S.B. Ko are with the Department of Electrical and Computer Engineering, University of Saskatchewan, Saskatoon, SK, S7N 5A9, Canada (e-mail: {s.m.mazhari, c.y.chung, seokbum.ko}@usask.ca).

temperature into account. In this respect, the process is accomplished by employing weather data and a heat balance equation. In contrast, measured values, such as sag position, mechanical tension, etc., are used in indirect DTLR calculation. In hybrid approaches, the indirect DTLR calculation, along with weather-related factors, are employed. Direct approaches are currently of notable interest as they are less dependent on external equipment and lead to low-cost outcomes compared to their counterparts [7].

From the prediction output point of view, DTLR prediction models can be divided into deterministic [18]-[22] and probabilistic [12]-[17] categories. In deterministic prediction, a single quantity associated with the most likely value of DTLR in the next sample point is considered as the output; in the probabilistic approach, information about uncertainty in DTLR prediction can be acquired. Because any overestimation of DTLR may lead to unprecedented issues, the probabilistic prediction is most welcome by power system operators [5] and accordingly is the focus of this work. The recommendations of IEEE and CIGRE joint task force (JTF) [4], further emphasizes the importance of probabilistic DTLR prediction. According to JTF's recommendation, the OHL's conductor temperature should be less than the maximum allowable conductor temperature for 99% of the time [4]. Therefore, it is critical to assess the performance of the DTLR prediction models for extreme probability of exceedance (POE) values (e.g., 99%); however, this is not well-discussed in the recent literature (i.e., [12], [15]-[17]).

In [12], [13], parametric distributions model the uncertainties associated with climatology variables. In [14], Taylor series expansion is employed to find the mean and variance of DTLR in the coming hours on the basis of the forecasted values of mean and variance corresponding to the meteorological variables. To consider the interdependency among meteorological variables in DTLR forecasting, in [26], multivariate Gaussian distributions, resulting from different meteorological variables, are used in a Monto Carlo simulation process to extract the DTLR distribution.

In [12]-[24], [25], a parametric representation is assumed for the uncertainties associated with meteorological variables. However, this assumption may be erroneous due to the high nonstationarity of the meteorological time series (TS). To this end, the authors of [15] propose a non-parametric autoregression framework for probabilistic DTLR prediction based on quantile regression (QR). The authors of [5], [16] put forward a non-parametric DTLR prediction based on quantile regression forest (QRF), in which meteorological measurements and numerical weather predictions (NWPs) compose the input features of the prediction model. Overall, studies on non-parametric DTLR prediction are still in their infancy; thus, this paper is devoted to further contributing to this literature.

Weather-based DTLR prediction models mainly limit their inputs to those features obtained directly from meteorological measurements. However, from the DTLR formulations proposed in [2]-[3], the relations of meteorological variables with DTLR value are evidently both complex and nonlinear. In addition, many latent variables (e.g., convection cooling, radiated heat loss rate, etc.) are acquired in the process of DTLR calculation and may provide information about the complex relationship between the DTLR value and meteorological variables. However, to the best of the authors' knowledge, these important predictors have thus far not been considered in DTLR forecasting. Hence, this paper scrutinizes the impact of latent variables in DTLR forecasting.

The remarkable advancements in deep learning and its successful implementation in prediction result in forecasting accuracy enhancement for a range of power systems applications [27]-[30]. In deep network (DNN) architectures, highly efficient unsupervised dimension reduction blocks (i.e., autoencoder variants) can be employed to tackle the high dimension feature space issue, posed due to numerous meteorological and latent variables [30]-[31]. Using DNNs, more complex patterns, which cannot be identified in shallow networks, can be perceived. Compared to various building blocks of DNNs, long short-term memory (LSTM) [32]—a recurrent NN (RNN)—has demonstrated superior performance in meteorological variable prediction problems [28]. The benefit of LSTM compared to conventional RNNs is its ability to capture long- and short-term dependencies in a sequence while addressing the vanishing and exploding gradient problems of prevalent RNNs [33]. Despite successful applications of RNNs, specifically LSTM, and the benefits of DNNs, these approaches have not been adapted to DTLR prediction. In this paper, LSTM is employed to develop the DTLR prediction, while the stacked denoising autoencoder (SDAE) [31] is developed for unsupervised feature learning and extraction. Most of the DNN models developed thus far have been trained deterministically in power systems applications [27]-[30]). A major impediment that made past methods incapable of offering a probabilistic model in DNNs may originate from the lack of a proficient probabilistic cost function; such function should consider reliability level and sharpness together so that the DNN parameters are tuned according to the system operator's preference. To address this need, inspired by the optimization function in

QR problem, a cost function is presented. The proposed cost function can be used to train the model for the preferred POEs.

This paper proposes a DTLR prediction for predicting the various POEs using accessible latent variables in addition to meteorological measurements. A DTLR model is developed using the SDAE and LSTM unit in the DNN architecture. The prediction engine is trained by considering a novel cost function to meet the JTF recommendation while the sharpness is maximized. The performance of the proposed models is compared with state-of-the-art DTLR prediction models using publicly available data. Briefly, the main contributions of this work are three-fold:

- For the first time, latent variables are introduced in DTLR forecasting as valuable predictors;
- A cost function is proposed to train the deep-learning model for probabilistic forecast;
- A deep learning model is trained for DTLR prediction.

## II. PROPOSED DYNAMIC THERMAL LINE PREDICTION FRAMEWORK

Hereafter, $\{WS_t\}_{t=1}^N$ and $\{WD_t\}_{t=1}^N$ are TS associated with wind speed and wind direction, respectively. $N$ is the length of the TS, and index $t$ refers to the $t$th sample of the TS. Also, the TS of the wind speed components, decomposed by a Cartesian coordinate system, are indicated by $\{WSX_t\}_{t=1}^N$ and $\{WSY_t\}_{t=1}^N$. The TS of ambient temperature and solar irradiance are represented by $\{AT_t\}_{t=1}^N$ and $\{SI_t\}_{t=1}^N$, respectively.

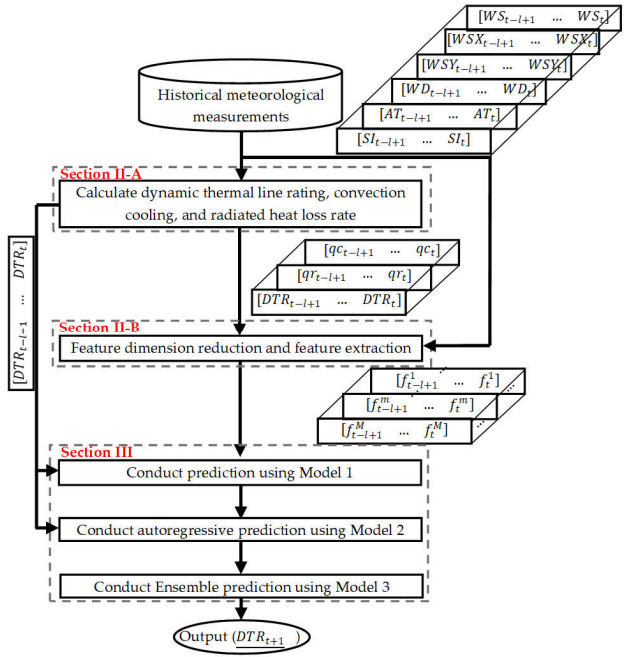

Fig. 1. General scheme of the proposed DTLR prediction.

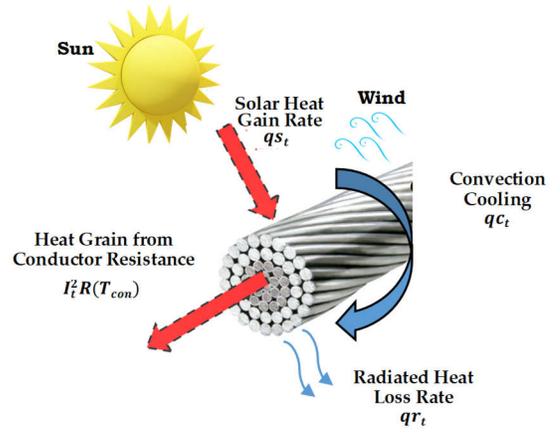

Fig. 2. A schematic diagram of DTLR.

Figure 1 demonstrates the overall framework of the proposed DTLR prediction model. The suitable number of lags ($l$) to form the input vector is identified using Cao's embedding dimension approach [34]. From the measurements, $l$ lags of meteorological TS are imported to the DTLR calculation model described in Section II-A. Using the DTLR calculation model, $[DTR_{t-l+1} \; \cdots \; DTR_t]$, $[qc_{t-l+1} \; \cdots \; qc_t]$, and $[qr_{t-l+1} \; \cdots \; qr_t]$, which respectively represent DTLR, convection cooling, and radiated heat loss rate for lags of meteorological inputs, are calculated. The features obtained from this step along with the meteorological input vector are utilized in a feature reduction and extraction stage as discussed in Section II-B. Thereafter, various trained models are employed to acquire the final DTLR prediction ($\underline{DTR_{t+1}}$) as elaborated in Section III.

### A. DTLR Calculation

In this study, the power system is assumed to be under normal operation; therefore, the current fluctuations and the resulting variations in the OHLs temperature can be considered negligible, provided the system does not require any abrupt and temporary switching [21]. Studies reveal that the maximum time required to reach the steady state because of a step change in current is approximately 30 minutes [35]. Based on these points, the transients in DTLR can be ignored, and DTLR can be estimated for every hour.

Figure 2 schematically represents the factors influencing the DTLR of OHLs. As per IEEE Std 738-2012, in the steady state, the heat balance equation for an OHL at the $t$th sample can be written as follows [2]:

$$qc_t + qr_t = qs_t + I_t^2 \cdot R(T_{con}) \tag{1}$$

where $qc_t$ and $qr_t$ are convection cooling and radiated heat loss rates per unit length, respectively. $qs_t$ in (1) is heat gain rate from the sun, and $R(T_{con})$ is alternating current (AC) resistance associated with the conductor temperature, $T_{con}$. From (1), the allowable conductor current, $I_t$, at $T_{con}$ can be simply obtained as follows:

$$I_t = \sqrt{\frac{qc_t + qr_t - qs_t}{R(T_{con})}} \tag{2}$$

From (1)–(2), one can observe that $qc_t$ and $qr_t$ are the cooling elements in the heat-balance equation, and their increase can help to obtain more OHL ampacity, while $qs_t$ is a heating component and is a culprit of ampacity reduction. $qc_t$ is a function of $WS_t$, $WD_t$, and $AT_t$. $qc_t$ is calculated as follows [2]:

$$qc_t = \max(qc_t^{f1}, qc_t^{f2}, qc_t^n) \tag{3}$$

$$qc_t^{f1} = K_a \cdot \left[1.01 + 1.35 \cdot N_{Re}^t \,^{0.52}\right] \cdot k_f \cdot (T_{con} - AT_t) \tag{4}$$

$$qc_t^{f2} = K_a \cdot 0.754 \cdot N_{Re}^t \,^{0.6} \cdot k_f \cdot (T_{con} - AT_t) \tag{5}$$

$$qc_t^n = 3.645 \cdot \rho_f^{0.5} \cdot D_o^{0.75} \cdot (T_{con} - AT_t) \tag{6}$$

$$N_{Re}^t = \frac{D_o \cdot \rho_f \cdot WS_t}{\mu_f} \tag{7}$$

where $K_a$ is wind direction factor; $\rho_f$, $\mu_f$, $k_f$ are air density, absolute air viscosity, and coefficient of thermal conductivity of air, respectively; and $D_o$ is the outside diameter of the conductor. As can be observed from (3)–(7), $qc_t$ is a nonlinear and complicated function of meteorological variables while the relationship between $qc_t$ and $I_t$ is simple as shown in (2). Therefore, considering $[qc_{t-l+1} \; \cdots \; qc_t]$ as elements of the predictor set for forecasting $DTLR_{t+1}$ can be beneficial. Moreover, $qr_t$ in (1)–(2) can be calculated as follows [2]:

$$qr_t = 17.8 \cdot D_o \cdot \epsilon \cdot \left[\left(\frac{T_{con}+273}{100}\right)^4 - \left(\frac{AT_t+273}{100}\right)^4\right] \tag{8}$$

where $\epsilon$ is emissivity and has a value between 0.23 and 0.91, which increases with conductor age. As can be seen from (8), $qc_t$, $qr_t$ is related to the fourth power of $AT_t$; therefore, considering the simple vector of $[AT_{t-l+1} \; \cdots \; AT_t]$ as the elements of the feature set may not be adequate to reflect the importance of ambient temperature and radiative heat loss in DTLR prediction, and it is worthful to analyze the influence of considering historical data of radiated heat loss rate. For a solar irradiance at time $t$, $SI_t$, the rate of solar heat gain, $qs_t$, can be estimated by a linear function of $SI_t$ [2]; therefore, its consideration as a feature cannot be informative.

The calculated DTLR, $[DTR_{t-l+1} \quad ... \quad DTR_t]$, as well as the latent variables obtained during the DTLR calculation, are used in a feature reduction and feature learning stage as elucidated below.

### B. Feature Reduction and Extraction in DTLR Prediction

As detailed in Section II-A, several climatic variables strongly influence DTLR value, including wind speed, wind direction, wind speed Cartesian components, ambient temperature, and solar irradiance. A tensor, formed by a series of lags associated with these variables, contains the potential informative predictors for DTLR. Moreover, the latent variables (i.e., convection cooling and radiated heat loss rate) may also contain valuable information about the complex relationship between climatic variables and DTLR. Historical DTLR values could also contain useful information.

#### 1) Feature Reduction

Properly optimizing the input features of the prediction engine by eliminating the non-informative and redundant features and identifying the features, which can demonstrate the DTLR pattern more efficiently, is a principal stage in DTLR prediction. To this end, we first employ a feature reduction stage based on minimal-redundancy-maximal-relevance (mRMR) [36]. mRMR is a mutual information ($MI$)-based approach that is employed in various power system problems to identify the subset of features providing the most information about the observation (target variable) [37]. $MI$ is widely used in the feature selection literature to evaluate the degree of uncertainty that a predictor can alleviate from an observation by measuring the mutual relevancy of predictor and target variable. For two random variables with domains $A$ and $B$, the $MI$ is defined as follows [33]:

$$MI(A;B) = \sum_{a \in A} \sum_{b \in B} P(a,b) \cdot \log\left(\frac{P(a,b)}{P(a) \cdot P(b)}\right) \quad (9)$$

where $P(a,b)$ represents the joint probability density function, and $P(a)$ and $P(b)$ are individual probability density functions of $a$ and $b$ random variables, which are the discretized format of continuous predictor and target variables. Based on (9), the iterative mRMR algorithm is carried out by the following optimization problem [33]:

$$\max_{\substack{a_j^{t-p} \in \mathcal{A} - \mathbf{\Omega_{n-1}}, \\ j=\{1,...,9\}, p=\{0,...,l-1\}}} \left[ MI(a_j^{t-p}; \mathbf{DTR_{t+1}}) - \right. \quad (10)$$

$$\left. \frac{1}{n-1} \sum_{\substack{a_i^{t-q} \in \mathbf{\Omega_{n-1}} \\ i=\{1,...,9\}, q=\{0,...,l-1\}}} MI(a_j^{t-p}; a_i^{t-q}) \right.$$

$$\mathcal{A} = \{a_1^{t-l+1} = \mathbf{WS_{t-l+1}}, ..., a_1^t = \mathbf{WS_t},$$
$$a_2^{t-l+1} = \mathbf{WD_{t-l+1}}, ..., a_2^t = \mathbf{WD_t},$$
$$a_3^{t-l+1} = \mathbf{AT_{t-l+1}}, ..., a_3^t = \mathbf{AT_t},$$
$$a_4^{t-l+1} = \mathbf{SI_{t-l+1}}, ...,$$
$$a_4^t = \mathbf{SI_t}, a_5^{t-l+1} = \mathbf{WSX_{t-l+1}}, ..., a_5^t = \mathbf{WSX_t}, \quad (11)$$
$$a_6^{t-l+1} = \mathbf{WSX_{t-l+1}}, ..., a_6^t = \mathbf{WSX_t},$$
$$a_7^{t-l+1} = \mathbf{qc_{t-l+1}}, ..., a_7^t = \mathbf{qc_t},$$
$$a_8^{t-l+1} = \mathbf{qr_{t-l+1}}, ..., a_8^t = \mathbf{qr_t},$$
$$a_9^{t-l+1} = \mathbf{DTR_{t-l+1}}, ..., a_{13}^t = \mathbf{DTR_t}\}$$

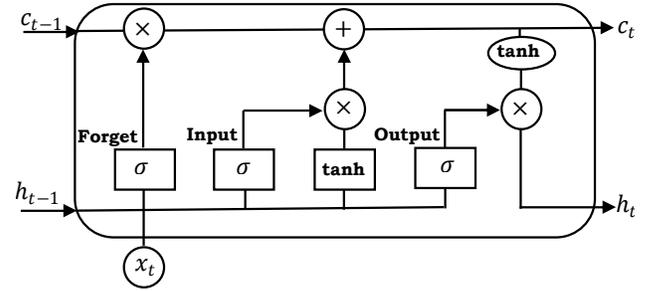

Fig. 3. The schematic of the LSTM unit.

where $\mathbf{\Omega_{n-1}}$ is the subset containing selected features at the $n-1th$ iteration. Bold notation represents the random variables of various predictors. $a_j^{t-p}$ is the random variable describing the $p$th lag of the $j$th feature. $\mathcal{A}$ is a set consisting of the random variables associated with feature candidates. Bold notation in the set $\mathcal{A}$, represents the vector related to feature candidates. In the first iteration of solving (10), $n=1$ and $\mathbf{\Omega_0} = \emptyset$ where $\emptyset$ refers to an empty set. In this work, the optimization problem in (10) is iteratively solved until $MI$ of the last component selected from solving (10) with $\mathbf{DTR_{t+1}}$ is negligible.

After conducting mRMR, the feature types that are not among the selected features in $\mathbf{\Omega_n}$ or constitute a negligible portion of $\mathbf{\Omega_n}$ are removed from the feature pool. The selected features, along with their corresponding $l$ lags, are then used to build the tensor input of the deep learning-based feature extraction described in Section II-B3.

#### 2) LSTM

LSTM is a unit of RNNs in which the temporal dependency among elements of the TS can be captured. An LSTM block consists of a memory cell, an input gate, an output gate, and a forgetting gate. The memory cell stores values for arbitrary time intervals. In LSTM, the three gates are neurons with activation functions. Figure 3 represents an LSTM unit. In this figure, $c_{t-1}$ and $h_{t-1}$ respectively denote the cell memory state and hidden state at the previous time. The input vector ($x_t$), $c_{t-1}$, and $h_{t-1}$ are used to update the memory state, $c_t$, and attain the output, $h_t$, corresponding to $x_t$. In Fig. 3, blocks labelled by $\sigma$ refers to sigmoid layers. The LSTM has been used in the proposed DTLR prediction as described in the next sections.

#### 3) RNN-based SDAE

Autoencoder (AE) is a type of neural network mainly employed for unsupervised feature learning in a wide range of applications [30], [31]. An AE is composed of two fragments: the encoder and decoder. The input of the encoder is the original feature tensor, which is mapped to a different space in the output of the encoder, while the decoder uses the mapped features as the input and reconstructs the original feature space. The output of the encoder part of the AE can be used as the input features of the following hidden layers. In the conventional AE, in case that the number of hidden layer nodes is higher than the input feature, the AE can potentially find the identity map. As a result, the AE can become inefficient. To circumvent this issue, denoising autoencoder (DAE) is proposed [30],[31]. In the DAE, the original input features are reconstructed from corrupted ones. In this way, the number of nodes in the DAE hidden layers can be larger than the number of input features without identity mapping risk;

therefore, this overcomplete DAE may disclose some informative features from the interactions among the input feature. In the following, the DAE is briefly described, while a detailed description of DAE can be found in [31].

At time $t$, all features that remain after the feature reduction stage, described in Section II-B1, are used to construct the input vector of the DAE, $x_t$, a three-dimensional tensor ($1 \times l \times m$) where $m$ is the number of feature variants. To capture the sequential correlation of the TS, LSTM, described in Section II-B2, are used as building blocks of the DAEs in this paper. An RNN-based DAE can be formulated as follows:

$$\min_{\theta, \theta'} (L_{DAE}(\boldsymbol{x}, \boldsymbol{z})) \tag{12}$$

$$\tilde{x} \sim q_D(x_t) \tag{13}$$

$$y_t = f_\theta(\tilde{x}_t, \boldsymbol{m}_{t-1}) \tag{14}$$

$$z_t = g_{\theta'}(y_t, \boldsymbol{m}'_{t-1}) \tag{15}$$

where $L_{DAE}(\cdot)$ is the loss function. In this paper, mean-squared-error (MSE) is employed as the loss function. The tensor $\boldsymbol{x}$ ($n \times l \times m$) contains all tensors $x_t$, $t = 1, \dots, n$, where $n$ is the number of available points in the validation set; tensor $\boldsymbol{z}$ consists of all tensors $z_t$, $t = 1, \dots, n$, which are the output of a DAE corresponding to input $\boldsymbol{x}$. $\theta$ and $\theta'$ encapsulate the unknown parameters of encoder and decoder blocks, respectively. Equation (13) represents the stochastic process of destroying some elements of input vector $x_t$ to form the corrupted input, $\tilde{x}$. In (14) and (15), $y_t$ and $z_t$ are the outputs of encoder and decoder blocks, respectively. $\boldsymbol{m}_{t-1}$ and $\boldsymbol{m}'_{t-1}$ are information passed from the calculation of $y_{t-1}$ and $z_{t-1}$, respectively, as the result of recurrent units. In (14) and (15), $f_\theta(\cdot)$ and $g_{\theta'}(\cdot)$ are the functions associated with encoder and decoder blocks, respectively.

Stacking several DAE forms an SDAE, in which more informative features can be extracted. Using the ADAM stochastic optimization [38], the SDAE is first trained layer-wise to find the initial parameters of all DAEs. Thereafter, and a fine-tuning process is conducted to further tune the SDAE, as elaborated in Section III. The output of the encoder portion of the SDAE is used as the input of Model 2 in the proposed DTLR prediction, as discussed in Section III.

## III. Proposed Training Framework for Probabilistic Deep Learning-based DTLR Prediction

Training of the proposed DTLR is conducted in several steps according to the scheme presented in Fig. 4. First, the data are divided into three parts: training, validation, and test datasets. The training dataset is used to tune the model parameters, while the validation dataset is used to assess the performance of the model during training. The test dataset is used to evaluate the trained model. The redundant or non-informative features are removed as described in Section II-B1. Next, the SDAE is formed and trained as discussed in Section II-B3. The autoregressive prediction model, named Model 1, is then trained using a series of DTLR TS lags; and simultaneously, a many-to-one prediction model, named Model 2, is trained using the feature tensor obtained from the encoder portion of the SDAE. Models 1 and 2 are both based on RNNs and trained using MSE as the cost function, while the weights of the SDAE are frozen. Then, the final model (Model 3) that employs the predictions of Models 1 and 2 as the input to provide a final prediction of the DTLR is trained deterministically using MSE

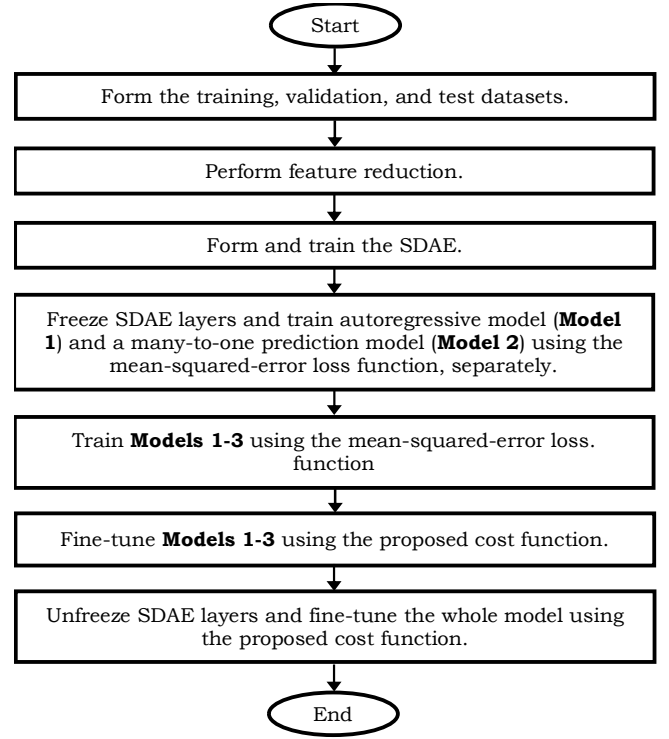

Fig. 4. General training scheme of the proposed probabilistic DTLR prediction.

as the cost function.

Afterwards, the Models 1-3 are further trained using the proposed cost function (16)-(20). Once Models 1-3 are tuned considering the proposed cost function, SDAE layers are unfrozen to tune the model further using the proposed cost function.

Inspired by QR formulation, the proposed cost function for training DNN for a preferred POE ($POE^*$) is defined as follows:

$$C\left(DTR_{t+1}, \underline{DTR_{t+1}}\right) = \sum_{i=1}^{i=2} a_i(sign(\epsilon_i) + 1) \cdot \epsilon_i \tag{16}$$

$$\epsilon_1 = \underline{DTR_{t+1}} - DTR_{t+1} \tag{17}$$

$$\epsilon_2 = DTR_{t+1} - \underline{DTR_{t+1}} \tag{18}$$

$$a_1 = POE^* \tag{19}$$

$$a_2 = 1 - POE^* \tag{20}$$

where $sign(\cdot)$ refers to the sign function, and $POE^*$ refers to preferred POE. In (16), the term associated with $\epsilon_1$ penalizes the prediction model when the prediction model results in $\underline{DTR_{t+1}}$ values above $DTR_{t+1}$. The more $\underline{DTR_{t+1}}$ value is higher than $DTR_{t+1}$, the more the cost function penalizes the model. On the other hand, when $\underline{DTR_{t+1}} < DTR_{t+1}$, the term related to $\epsilon_1$ in the cost function becomes zero; however, the term corresponding to $\epsilon_2$ penalizes the prediction model if $\underline{DTR_{t+1}}$ is deviated from the $DTR_{t+1}$.

Equation (16) is non-differentiable at $\epsilon_i = 0$ due to $sign(\cdot)$. This can cause an issue in the backpropagation process during the training phase. In this regard, $sign(\cdot)$ is approximated by $tanh(\cdot)$, and (16) can be rewritten as follows:

$$C\left(DTR_{t+1}, \underline{DTR_{t+1}}\right) = \sum_{i=1}^{i=2} a_i(tanh(C \cdot \epsilon_i) + 1) \cdot \epsilon_i \tag{21}$$

where $C$ is a large constant number so that $tanh(C \cdot \epsilon) \to 1$ when $\epsilon_i > 0$, while $tanh(C \cdot \epsilon_i) \to -1$ for $\epsilon_i < 0$. $tanh(C \cdot \epsilon_i) = 0$ when $\epsilon_i = 0$. Using $tanh(\cdot)$ results in a differentiable function that can use the off-the-shelf DNNs optimizers for training the model.

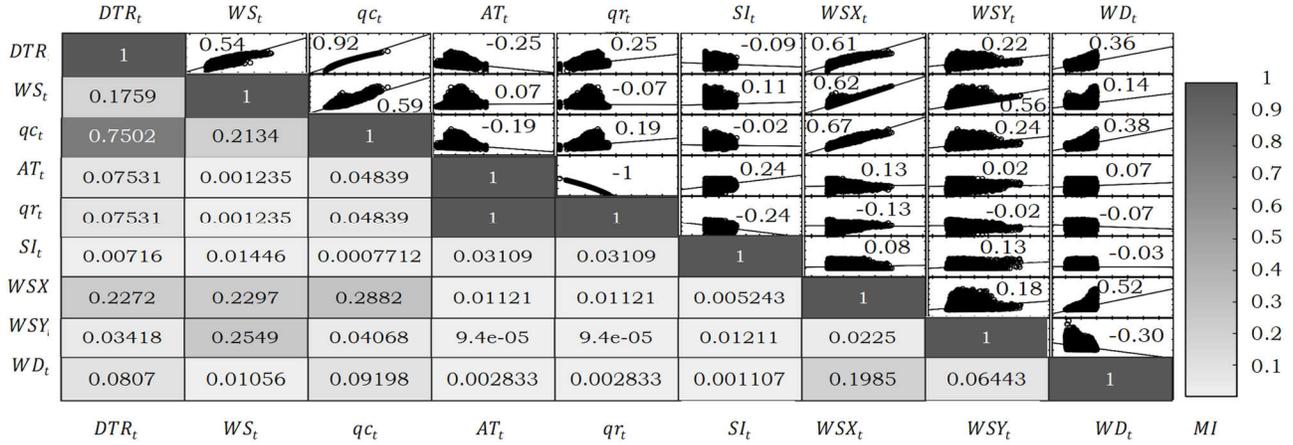

Fig. 5. Dispersion of DTLR with respect to different feature types and Kendall $\tau$ rank coefficient and mutual information of different features and DTLR.

The proposed cost function enables fine-tuning the DTLR model such that the DTLR prediction model provides the lower bound of DTLR values with minimum deviations from the actual DTLR values.

## IV. CASE STUDIES AND COMPARISONS

### A. Descriptions of Data, Simulation Tools, and the Proposed Model Settings

We performed the analysis based on a 5-year dataset (Jan. 1, 2010 to Jan. 1, 2015), recorded from the M2 met tower at the National Wind Energy Center (NWEC) [36], located in Denver, USA. Linear interpolation is employed to fill the missing points in the dataset. It is assumed that the measurements correspond to an OHL constructed from 400 $mm^2$ Drake 26/7 ACSR conductor at the elevation of 1861 $m$ from sea level.

Based on [2], the specification of this conductor is summarized in Table I, where the STLR is calculated based on the low-speed perpendicular wind (0.6 m/s), high ambient temperature (40 °C), and full solar heating (1000 W/m²) [40].

The proposed model was implemented in Python 3.7 on a Windows 10 PC with a 1.6 GHz Intel Core i5 CPU and 8 GB of memory. Based on the grid search, three DAEs are used, respectively, are selected as the suitable architecture for SDAE. While the number of hidden layers for Models 1-2 are one and three, respectively. While Model 3 is a single layer model which provides the final forecast by using the outputs from Models 1-2 The training is conducted using the process described in Section III.

### B. Analyzing Feature Candidates

To evaluate the dependency of $DTR$ on different feature candidates, a 5-year meteorological dataset is used to generate historical $DTR_t$, $qc_t$, and $qr_t$ TS, using the procedure described in Section II-A.

Figure 5 represents the dispersion of $DTR$ with respect to some of the feature types in the upper triangle.

The Kendall $\tau$ rank coefficient is widely employed as a non-parametric statistical test in hypothesis testing to identify the statistical dependency between two random variables [41].

TABLE I
SPECIFICATIONS OF THE CONDUCTOR

| | |
|---|---|
| STLR ($A$) | 685 |
| Outside diameter of conductor ($mm$) | 28.12 |
| Minimum conductor temperature (°C) | 25 |
| Maximum conductor temperature (°C) | 75 |
| Conductor resistance at minimum temperature ($\Omega/km$) | 0.07284 |
| Conductor resistance at maximum temperature ($\Omega/km$) | 0.08689 |
| Solar absorptivity and emissivity | 0.5 |

Detailed explanations of the Kendall $\tau$ rank coefficient can be found in [38]. The presented quantities in the subplots, forming the upper triangle, represent the Kendall $\tau$ rank coefficient of different features vis-à-vis each other and $DTR_t$. Based on the calculated Kendall $\tau$ rank coefficients, it can be concluded that there is a strong relation between $DTR_t$ and latent and the meteorological variables.

Furthermore, this figure shows that $qc_t$ is the most influential factor in $DTR_t$ compared to all other features. This strong relation can also be found from the $MI$ values presented in the diagonal subplots and lower triangle of Fig. 5. The $MI$ values testify to the high dependency of $DTR_t$ on $qc_t$. One can note from this figure that, among the meteorological variables, $WS$ demonstrates the strongest correlation with $DTR$ while $MI(WS_t; DTR_t)$ is significantly lower than $MI(qc_t; DTR_t)$. On the other hand, $qr_t$ and $AT_t$ are fully correlated based on Kendall $\tau$ rank coefficient as well as $MI$; therefore, considering only one of them could be adequate in DTLR prediction. Moreover, from $MI(DTR_t; SI_t)$, it can be recognized that $SI_t$ can provide the minimum information in DTLR forecasting. To this end, this feature species is eliminated in the feature reduction stage described in Section II-B. It is worth noting that the data are discretized with respect to median values of different features to calculate the $MI$ values. This type of discretization is widely used for calculating $MI$ among continuous variables [42].

### C. Description of Benchmark Models

Three benchmark models are utilized in this paper: persistence [12], QR [15], and QRF [16] prediction models. The persistence model (PM) is a conventional prediction model that is widely used for short-term prediction of meteorological-related variables including DTLR [12]. The simplicity of PM can facilitate the comparison of the proposed model with other prediction models. QR was recently proposed for the non-parametric probabilistic prediction of the DTLR, which is the focus of this study. Therefore, comparing the efficacy of the proposed model with QR provides some insights about the superiority of the proposed model with respect to other non-parametric probabilistic DTLR prediction models. As the last benchmark model, the state-of-the-art QRF-based probabilistic DTLR prediction model is employed [5], [16]. Optimal hyperparameters of the QRF is found using the Bayesian optimization. To carry out a fair comparison, the proposed model and benchmark models utilize similar input variables, as elaborated in Section II-A, but a notable exception is PM, for which the mean and variance of the latest DTLR values form a Gaussian distribution representing the uncertainty of upcoming samples.

### D. Evaluation Metrics

Three evaluation metrics are used to appraise the performance of the different DTLR models. The probability of exceedance (POE) of the prediction model is the most imperative evaluation criterion for a secure probabilistic DTLR prediction and is defined as follows:

$$POE(\%) = \frac{1}{N} \sum_{n=l}^{N} \delta_n \times 100 \qquad (22)$$

$$\delta_n = \begin{cases} 1 & \overline{DTR_{t+1}} \leq DTR_{t+1} \\ 0 & otherwise \end{cases} \qquad (23)$$

where $N$ is the number of points in the training, validation, or test datasets. Any deviation of the $POE$ from the preferred $POE$ ( $POE^*$), can lead to unprecedented issues. As a measure of sharp
ness of the predicted $POE$s, normalized mean absolute error ($NMAE$) is used which is defined as follows:

$$NMAE(\%) = \frac{\frac{1}{N} \sum_{t=1}^{N} \left| DTR_{t+1} - \overline{DTR_{t+1}} \right|}{R} \times 100 \qquad (24)$$

where $R$ is the range of DTLR values. In a perfect DTLR prediction, $POE = POE^*$ while $NMAE \rightarrow 0$ . Root-mean-squared-error (RMSE), which is a valuable measure and signifies large deviations of DTLR prediction from its actual value, is also employed as another evaluation metric. $RMSE$ can be calculated as follows:

$$RMSE = \sqrt{\frac{1}{N} \sum_{t=1}^{N} \left( DTR_{t+1} - \overline{DTR_{t+1}} \right)^2} \qquad (25)$$

### E. Numerical Studies

As an instance, the performance of the prediction model on a six-month dataset (Jan. 1, 2010 to Jan. 1, 2012) is used for numerical comparisons. Most (85%) of the data are used for training and validation, while the remainder (15%) are employed for testing. In this section, the effectiveness of considering latent predictors (i.e., convection cooling and radiated heat loss rate) is first investigated. Thereafter, the proposed DTLR is compared with the benchmark models for different $CL$ preferences, using various evaluation metrics.

#### 1) The Impact of Latent Predictors

Section IV-B showed that, in comparison to directly observed meteorological variables, the radiated heat loss rate has more a pretentious relation with DTLR.

To empirically investigate the efficacy of considering a series of lags associated with the mentioned latent variables as predictors, a case study is conducted in this section using the proposed DTLR prediction with and without the latent predictor. Table II summarizes the case study conducted for $POE^* = 99\%$ . The table shows the experiential results are in line with the theoretically expected outcome, and considering latent variables can reduce $NMAE$ and $RMSE$, while the $POE$ criterion is satisfied. Therefore, using the proposed prediction approach unlocks the OHL ampacity further, while the $POE$ is satisfied; as a result, more power can be transferred through the transmission systems. So, it can be concluded that the latent variables can provide further information about the DTLR pattern and, therefore, the prediction can be performed more precisely.

#### 2) Numerical Comparisons of Different Prediction Models

The performance evaluations of different prediction models for $POE^* = [90\%, 95\% , 99\%]$ are tabulated in Table III. The table shows that PM is incompetent in providing secure DTLR prediction for any of the $POE^*$ values. While other benchmark models along with the proposed model can fulfill the $POE^*$ requirements. From $NMAE$ and $RMSE$ viewpoints, for $POE^*$=90%, the performance of QR, QRF, and the proposed model are closely comparable. For $POE^* = 95\%$, based on $NMAE$ and $RMSE$, the proposed model slightly outperforms benchmark models while $POE^*$ is satisfied (i.e., $POE \geq POE^*$). For $POE^* = 99\%$, the case study demonstrates that the proposed model can facilitate secure DTLR employment, while $NMAE$ and $RMSE$ is reduced compared to the benchmark models. From this table, one can note that for $POE^* = 99\%$, the proposed model results in reductions in $NMAE$ and $RMSE$ by at least 6.33% ($\frac{18.51 - 19.76}{19.76} \times 100$) and 5.78% ($\frac{309.16 - 291.30}{309.16} \times 100$), respectively.

TABLE II
COMPARING THE PERFORMANCE OF THE PROPOSED MODEL WITH AND WITHOUT LATENT VARIABLES AS INPUT

|  | **POE(%)** | **NMAE(%)** | **RMSE** |
|---|---|---|---|
| Without latent variable | 98.95 | 19.21 | 300.20 |
| With latent variable | 99.01 | 18.51 | 291.30 |

TABLE III
PERFORMANCE EVALUATION OF DIFFERENT PREDICTION MODELS

|  | **Prediction Models** | **POE (%)** | **NMAE (%)** | **RMSE** |
|---|---|---|---|---|
| $POE^* = 90\%$ | PM | 83.42 | 13.80 | 251.10 |
| | QR | 92.20 | 11.9 | 202.55 |
| | QRF | 92.64 | 12.0 | 203.38 |
| | Proposed Model | 91.90 | 11.91 | 203.60 |
| $POE^* = 95\%$ | PM | 89.57 | 16.4 | 290.41 |
| | QR | 96.60 | 15.01 | 242.21 |
| | QRF | 95.56 | 14.73 | 241.27 |
| | Proposed Model | 95.14 | 14.50 | 236.61 |
| $POE^* = 99\%$ | PM | 95.75 | 22.12 | 372.52 |
| | QR | 99.80 | 19.96 | 311.80 |
| | QRF | 99.31 | 19.76 | 309.16 |
| | Proposed Model | 99.01 | 18.51 | 291.30 |

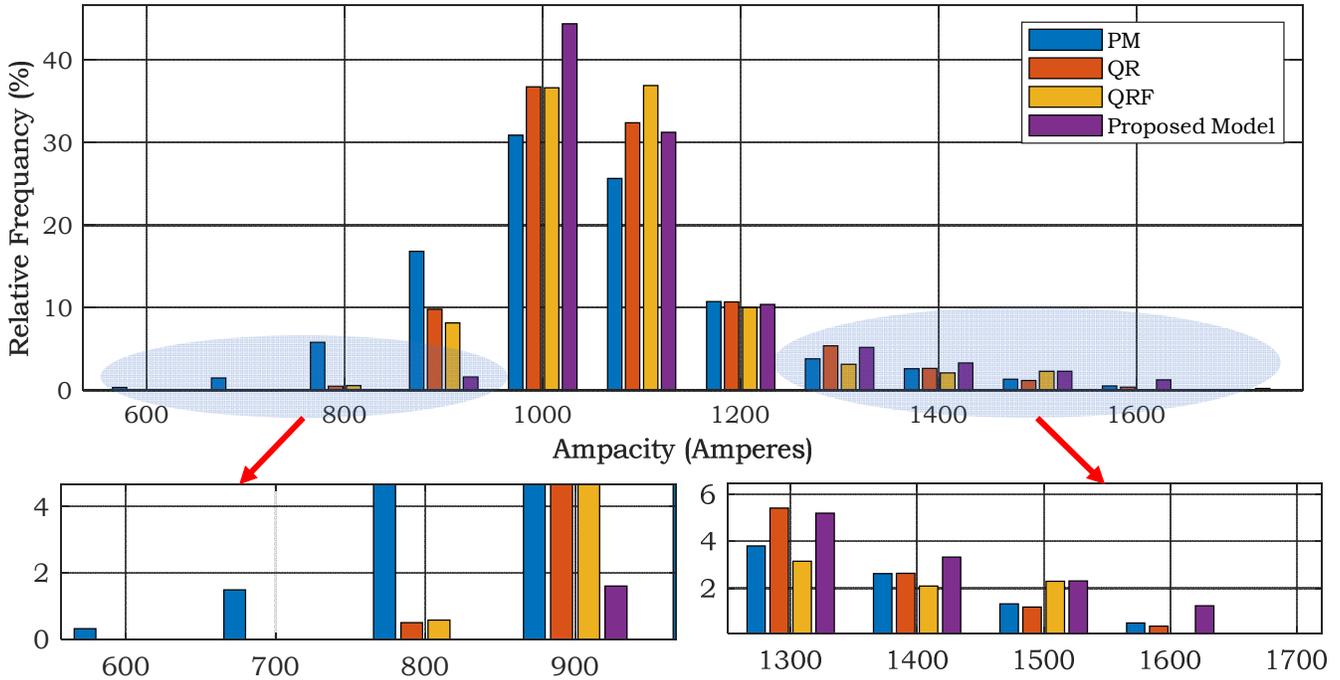

Fig. 6. Distribution of forecasted DTLR for $POE^* = 99\%$ with forecast values equal or less than the maximum actual ampacity.

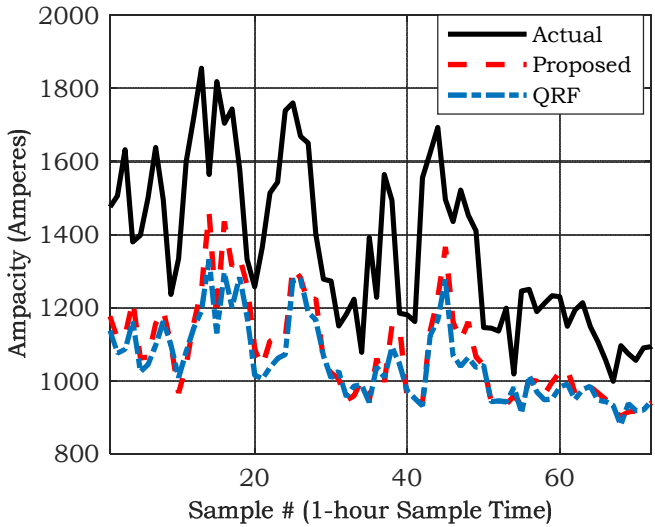

Fig. 7. Prediction of DTLR using the proposed model and QRF for $POE^* = 99\%$.

To further investigate the performance of the proposed model compared to benchmark models, the histograms of various DTLR prediction models for $POE^* = 99\%$ is presented in Fig. 6. It is worth noting that histograms are driven by considering only the prediction values that are equal to or less than the actual values. Thus, the performance of the different prediction models in terms of unlocking the OHL's capacity can be compared. As can be observed from this figure, for several points in the test dataset, the benchmark models result in ampacity predictions, which are equal or less than $900A$, while the proposed model leads to limited number $DTLR$ prediction values smaller than $900A$. On the other hand, for more than about 10% of the times the proposed model results in DTLR prediction which are greater than $1200A$, while the $POE^*$ criterion is met, while the frequency of DTLR prediction of higher than $1200A$ using the benchmark models is reduced.

Figure 7 depicts a sample episode for prediction for $POE^* = 99\%$ with the proposed DTLR and QRF—the best benchmark model for $POE^* = 99\%$. The figure covers the prediction for three successive days (i.e., 72 1-hour samples) to provide the view of the high volatility in DTLR values and efficacy of the proposed model to track this volatile pattern. As can be observed, the DTLR can vary substantially within an hour, which mainly stems from the chaotic behaviour of its influential factors (e.g., wind speed, wind direction, etc.). It is discerning from this figure that the proposed model can result in DTLR prediction values, which are $100A$ higher than the ones obtained from QRF. Besides, comparing the STLR, reported in Table I, with the DTLR forecasts shown in Figure 7, one can conclude that with minimal risk, a significant amount of OHL ampacity headroom can be unlocked.

## V. DISCUSSION

As case studies elucidate, the proposed DTLR prediction framework results in a more accurate and reliable forecast compared to various well-established benchmark models. In Sections IV-E1 and IV-E2, the detailed analyses demonstrate the superior performance of the proposed model results from both considerations of latent variables and the deep learning architecture. The proposed model can be used as a decision-support tool for system operators for unlocking the OHL's additional ampacity considering a pre-defined risk exposure acceptance (i.e., $POE^*$).

The hour-ahead secure DTLR prediction can facilitate renewable generation accommodation and alleviate the transmission congestion issues caused by renewables generation. Consequently, the wind power curtailment is reduced, while the need for transmission expansion can be postponed or even eliminated. Such a highly accurate hour-ahead forecast can be implemented in the energy management system for real-time security constraint economic dispatch [43].

In this study, the efficacy of the proposed model is validated for a range of $POE^*$ values. In practice, risk-based assessment of the transmission lines and the cost implications of the OHL's annealing— as a consequence of operating the OHL higher than the conductor temperature limit—can be used to identify the optimal $POE^*$[44].

The feature selection and extraction play crucial roles in the forecasting frameworks. In this paper, mRMR and SDAE, which are well-known feature selection and extraction approaches, are adapted, respectively. Despite the extensive research on the development of the advanced feature selection and extraction techniques in various power system related forecasting applications [45], the development of these machine learning building blocks in the DTLR literature is scarce. To this end, extensive research on feature selection and extraction in DTLR forecasting can be considered as an important future research direction.

The proposed DTLR forecasting framework is validated for a direct DTLR forecasting application in which only meteorological variables and the DTLR formulations are employed; however, considering the generality of the framework, as a future study, other features such as OHL's sag/tension as well as NWP can be incorporated in the framework to improve the performance of DTLR forecasting further. Furthermore, using the NWP can also facilitate extending the forecasting horizon, which is beneficial for day-ahead unit commitment and economic dispatch.

## VI. Conclusion

This paper proposed a deep learning-based probabilistic DTLR prediction model for hour-ahead power system operation problems. The latent variables, obtained in the process of calculating the DTLR value, are considered as new predictors of the proposed model, while SDAE is employed for feature learning and extraction. A training strategy is devised to train the DNN, and a cost function is put forward to train the prediction model in a probabilistic manner. The proposed prediction framework considers no hypothesis about the uncertainty of the DTLR. Simulation results confirm the efficacy of the proposed model and its superiority compared to benchmark models.

**N. Safari** received this B.Sc. (Hon.) degree from the K. N. Toosi University of Technology, Tehran, Iran, in 2012, the M.Sc. degree from the Tehran Polytechnic University, Tehran, Iran, in 2014, and the Ph.D. degree in electrical engineering from the University of Saskatchewan, Saskatoon, SK, Canada, in 2018. He is currently a Grid Operations Support Engineer-in-Training with SaskPower, Regina, SK, Canada. His research interests include applications machine learning algorithms in the integration of large-scale renewable energy generation to power systems. Dr. Safari has served as a Reviewer for several reputable journals including "IEEE Transactions on Power Systems", "IEEE Transactions on Sustainable Energy", "Applied Energy", "Applied Mathematical Modelling", and "Applied Soft Computing".

**S.M. Mazhari** received the M.Sc. (Hon.) degree from the University of Tehran, Tehran, Iran, in 2012, and the Ph.D. (Hon.) degree from the Amirkabir University of Technology, Tehran, Iran, in 2016, all in electrical engineering. He is currently a postdoctoral research fellow at the University of Saskatchewan, Saskatoon, SK, Canada.

**C.Y. Chung** received the B.Eng. (with First Class Honors) and Ph.D. degrees in electrical engineering from The Hong Kong Polytechnic University, Hong Kong, China, in 1995 and 1999, respectively.

He is currently a Professor, the NSERC/SaskPower (Senior) Industrial Research Chair in Smart Grid Technologies, and the SaskPower Chair in Power Systems Engineering in the Department of Electrical and Computer Engineering at the University of Saskatchewan, Saskatoon, SK, Canada. His research interests include smart grid technologies, renewable energy, power system stability/control, planning and operation, computational intelligence applications, power markets and electric vehicle charging.

Prof. Chung is a Senior Editor of "IEEE Transactions on Power Systems", Vice Editor-in-Chief of "Journal of Modern Power Systems and Clean Energy", Subject Editor of "IET Generation, Transmission & Distribution", Editor of "IEEE Transactions on Sustainable Energy" and "IEEE Power Engineering Letters", and Editorial Board Member of "CSEE Journal of Power and Energy Systems" and "Protection and Control of Modern Power Systems". He is also an IEEE PES Distinguished Lecturer and a member of IEEE PES Fellows Evaluation Committee.

**S.B. Ko** received his Ph.D. in Electrical and Computer Engineering at the University of Rhode Island, Kingston, Rhode Island, USA in 2002. He is currently a Professor in the Department of Electrical and Computer Engineering at the University of Saskatchewan, Saskatoon, Canada. He worked as a member of technical staff for Korea Telecom Research and Development Group, Korea, from 1993 to 1998. Dr. Ko is an associate editor for "IEEE Access" and "IEEE Transactions on Circuits and Systems I".